\documentclass[fleqn,usenatbib]{mnras}

\usepackage{newtxtext,newtxmath}

\usepackage[T1]{fontenc}
\usepackage{ae,aecompl}

\usepackage{graphicx}	

\usepackage{amsmath}	
\usepackage{amssymb}	
\usepackage{caption}
\usepackage{multicol}
\usepackage{etoolbox}

\title[Revisiting the Cygnus OB associations]{Revisiting the Cygnus OB associations}
\author[A. L. Quintana and N.J. Wright]{
Alexis L. Quintana$^{1}$\thanks{E-mail: a.l.p.quintana.isasi@keele.ac.uk} and
Nicholas J. Wright$^{1}$ \\
$^{1}$Astrophysics Group, Keele University, Keele ST5 5BG, UK\\
}

\date{Accepted 2021 September 13. Received 2021 September 13; in original form 2021 August 5}

\pubyear{2021}

\begin{document}
\label{firstpage}
\pagerange{\pageref{firstpage}--\pageref{lastpage}}
\maketitle

\begin{abstract}
OB associations play an important role in Galactic evolution, though their origins and dynamics remain poorly studied, with only a small number of systems analysed in detail. In this paper we revisit the existence and membership of the Cygnus OB associations. We find that of the historical OB associations only Cyg OB2 and OB3 stand out as real groups. We search for new OB stars using a combination of photometry, astrometry, evolutionary models and an SED fitting process, identifying 4680 probable OB stars with a reliability of $>$90\%. From this sample we search for OB associations using a new and flexible clustering technique, identifying 6 new OB associations. Two of these are similar to the associations Cyg OB2 and OB3, though the others bear no relationship to any existing systems. We characterize the properties of the new associations, including their velocity dispersions and total stellar masses, all of which are consistent with typical values for OB associations. We search for evidence of expansion and find that all are expanding, albeit anistropically, with stronger and more significant expansion in the direction of Galactic longitude. We also identify two large-scale (160 pc and 25 km s$^{-1}$) kinematic expansion patterns across the Cygnus region, each including three of our new associations, and attribute this to the effects of feedback from a previous generation of stars. This work highlights the need to revisit the existence and membership of the historical OB associations, if they are to be used to study their properties and dynamics.
\end{abstract}

\begin{keywords}
stars: kinematics and dynamics - stars: early-type - stars: massive - stars: distances - open clusters and associations: individual: Cyg OB1, Cyg OB2, Cyg OB3, Cyg OB8, Cyg OB9.
\end{keywords}



\section{Introduction}
Many OB stars are assembled in associations \citep{McKee}. They were first defined by \citet{Amb1947} as gravitationnally unbound groups of young stars with a high concentration of OB stars. They typically have total masses ranging from a few thousands to a few tens of thousands of stellar masses, with a density lower than 0.1 $M_{\odot}$ pc$^{-3}$ and a young age (necessary to explain their kinematic and spatial concentration despite their unbound nature). They often extend over tens of parsecs and include smaller groups of different ages and kinematics \citep{Garmany,Wright2020}. They have a huge effect on their environment: the feedback from their massive members creates H~{\sc ii} regions, superbubbles and dissociation in the diffuse interstellar medium \citep{Shull, Dove}. As a transitional phase between star-forming regions and the field population of stars, OB associations are important to study in order to better understand Galactic evolution \citep{Wright2020}.

The two main scenarios for the origin of OB associations are the expanding star cluster and hierarchical star formation models. According to the former, OB associations constitute the expanded remnants of dense star clusters \citep{Blauuw}, following a disruption of their parent molecular cloud through a feedback phase once massive stars emerge \citep{Hills}. According to the latter, stars form in groups with a wide variety of sizes and densities and most are unbound and disperse from birth \citep{Krui2012}. It is therefore considered important to search for expansion patterns in OB associations. This has been possible over the last few years thanks to data from various astrometric surveys, predominantly from the Gaia mission \citep{Gaia}. The results so far have been mixed, with \citet{Wright2016}, \citet{WrightMamajek} and \citet{Ward} failing to detect expansion, while \citet{Melnik2020} measured an expansion signature in only 6 of the 28  OB associations they studied. On the other hand, more recent studies that have used a revised kinematic selection of the members of OB associations have found expansion in several associations and their subgroups (\citealt{Kounkel2018},  \citealt{CantatGaudin2019}, \citealt{Armstrong2020}). Such findings encourage a deeper investigation.

The goal of this work is to study the kinematics of OB associations to search for evidence of expansion, using reliable and updated membership lists. This requires us to either refine the membership of existing OB associations or identify such systems from scratch. To that aim we turn our attention to the Cygnus region as a first target since it contains many OB associations and is rich in OB stars \citep{Reipurth}. 

The Cygnus region of the Galactic plane is home to a considerable amount of recent star formation, including the massive Cygnus X star forming region, thousands of OB stars, H~{\sc II} regions, supernova remnants and 9 OB associations \citep{Mel, Reipurth}. Since the study of \citet{Massey}, the highly-reddened OB star population of Cygnus OB2 has been studied by many authors, most recently by \citet{Wright} and \citet{Berlanas2020}. The association is home to approximately 70 O-type stars with ages between 1 and 7 Myr and a total mass of 1.65 x 10$^4$ $M_{\odot}$ \citep{Wright}. However, \citet{Berlanas2} recently showed that the association appears to constitute two structures super-imposed along the line of sight: the main association at a distance of $\sim$1.76 kpc and a smaller, foreground population at $\sim$1.35 kpc. Studies of the kinematics of Cyg OB2 have to date failed to identify a clear expansion signature \citep{Wright2016, Arnold}. The other OB associations in Cygnus remain poorly studied despite the available information (see e.g. \citealt{Reipurth, Comeron, Melnik2017, Melnik2020}) and deserve as a consequence a better understanding.

In this paper we investigate the 5 main OB associations in Cygnus, consider their reality as genuine associations, and eventually propose the existence of 6 alternative structures in the region, some of which overlap with the historical associations. In Section \ref{Known} we analyse the historical OB associations and show that, apart from Cyg OB2 and OB3, they do not stand out as real groups. In Section \ref{method} we describe how we identify OB stars across the region using photometry, astrometry, evolutionary models and a new SED fitting tool. In Section \ref{newassoc} we use a new flexible clustering technique to identify new associations. In Section \ref{analysis} we characterize the identified groups by studying their physical and kinematic properties. In Section \ref{discussion} we discuss these findings and in Section \ref{conclusions}, we wrap up our results. 
 
\section{The existing OB associations in Cygnus}
\label{Known}

In this section we consider the stellar content and reality of the existing OB associations in Cygnus using modern photometry and astrometry.

Our focus is on a 60 deg$^2$ area in the Cygnus constellation spanned by l = [71$^{\circ}$,81$^{\circ}$] and b = [-1$^{\circ}$, 5$^{\circ}$]. This region contains the OB associations Cyg OB1, OB2, OB3, OB8 and OB9, the five most well-studied associations in Cygnus. A total of 178 OB stars were identified by \citet{Blaha} and divided between these five associations based only on their positions on the sky. The other OB associations in Cygnus, i.e. Cyg OB4, OB5, OB6 and OB7 are not considered here because they are much more extended and found at greater longitudes. For this work we use the list of 164 OB stars in Cyg OB1, OB3, OB8 and OB9 from \citet{Blaha}. For Cyg OB2 we use the updated census of OB stars from \citet{Berlanas2}. This gave us a total of 341 OB stars.

These sources were crossmatched with Gaia EDR3 data using a matching radius of 1 arcsec, with 329 sources matched successfully. This data was filtered by applying the recommended astrometric criterion from \citet{Lindegren2020}, that their renormalised unit weight error (RUWE) should be lower than 1.4. 269 sources fulfilled these conditions. Distances of sources with good astrometry were obtained from \citet{BailerEDR3}.

\begin{figure}
\begin{center}
\includegraphics[width = \columnwidth]{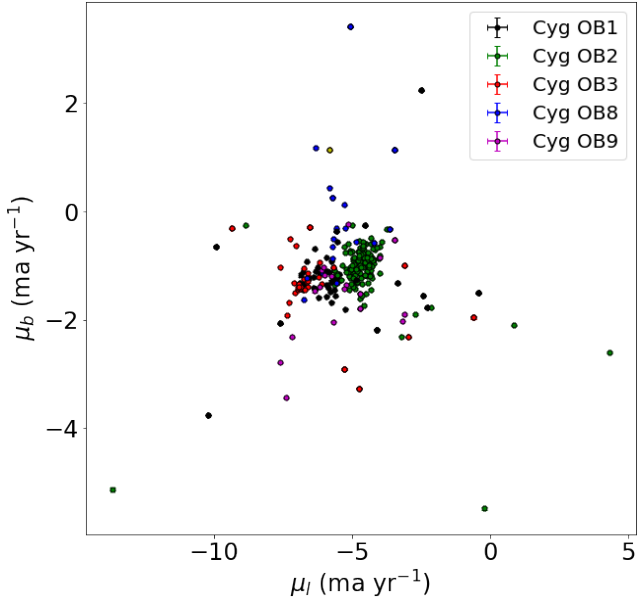}
\caption{Galactic proper motion distribution for the 121 objects from \citet{Blaha} and 148 objects from \citet{Berlanas2} that passed the astrometric quality cut, with error bars shown (the majority of which are smaller than the plotted symbol). It is clear that, with the exception of Cyg OB2 and OB3, the other associations are not kinematically distinct or coherent.}
\label{propermot}
\end{center}
\end{figure}

Figure \ref{SpaceDistib} shows the distribution of these sources in both Galactic coordinates (where all stars are shown) and with distance plotted against $l$ (where only sources that passed the astrometric quality tests are shown). Most associations are at similar distances, between 1.5 and 2.0 kpc, with a slight trend of increasing distance with decreasing Galactic longitude. Cyg OB2 is the most distinct and highly concentrated of the associations, with the other associations less concentrated, though all associations appear to include objects whose distances or kinematics (Figure~\ref{propermot}) would imply they are unlikely to be part of these associations.

\begin{figure*} 
    \centering
    \includegraphics[scale = 0.15]{SpaceDistrib.pdf}
    \caption{Top panel: spatial distribution of Cygnus OB association members from \citet{Blaha} and \citet{Berlanas2}, displayed with their proper motion as a vector, with a background extinction map from \citet{Bayestar}, representing the integrated reddening up to a distance of 1.8 kpc. Objects with $RUWE > 1.4$ have been displayed as small circles (whose proper motions may be questionable). Bottom panel: distance as a function of Galactic longitude for sources with $RUWE < 1.4$, with distances and uncertainties taken from \citet{BailerEDR3}. \label{SpaceDistib}}
\end{figure*}

We attempted to refine the association membership, as originally defined by \citet{Blaha}, by removing outliers in distance and proper motion, and also used a number of modern clustering algorithms applied to this list of OB stars to "re-discover" these associations, but only Cyg OB2 and OB3 appeared to show any level of kinematic coherence that would suggest they are true associations. We can only conclude that the other associations, Cyg OB1, OB8 and OB9, are not genuine OB associations. It would therefore be pertinent to reconsider the possible OB associations in Cygnus, starting from the ground up.

\section{Identifying and characterising OB stars}
\label{method}

In this section we outline our photometric SED fitting technique used to identify OB stars across our region of study and estimate their physical parameters.

\subsection{Data and selection process}
\label{Data}

We use photometry and astrometry from IGAPS (INT Galactic Plane Survey, providing $g$, $r$, $i$ band photometry, \citealt{Drew}, \citealt{Mongui}), 2MASS (2 Micron All Sky Survey, providing $J$, $H$, and $K_s$ band photometry, \citealt{2MASS}), UKIDSS (United Kingdom Infrared Deep Sky Survey, providing deeper $J$, $H$ and $K$ band photometry, \citealt{UKIDSS}), and Gaia EDR3 (\citealt{Gaia}, \citealt{GaiaEDR3}, \citealt{Riello}), providing $G$, $G_{BP}$ and $G_{RP}$ band photometry. In addition we use parallaxes and proper motions from Gaia EDR3 \citep{Fabricius}, and the probabilistic distances derived from these parallaxes provided by \citet{BailerEDR3}. We correct all parallaxes for the non-zero parallax zero-point, whose value has been shown to depend on magnitude, ecliptic latitude, and the pseudocolour of the source, using the prescription in \citet{ZeroPoint}.

There are a total of 8,474,415 sources in Gaia EDR3 in this region of study, 7,434,799 of which have a valid five or six-parameter astrometric solution (suitable for our needs). We crossmatch this sample with 2MASS, UKIDSS and IGAPS, using a matching radius of 1", using the proper motions to account for the epoch difference. We found near-IR matches for 6,347,730 sources and IGAPS matches for 6,158,261 sources. We discarded the 1,087,069 Gaia sources that lacked a near-IR counterpart as a near-IR photometric datapoint is necessary for our SED fitting. These sources generally have $G > 18$, too faint to be detected by 2MASS and lying in one of the areas of incomplete coverage for UKIDSS. They are unlikely to correspond to OB stars in our region of interest.

We discarded all unreliable photometry. For IGAPS, photometry is reliable if saturation is avoided and if its associated class indicates a star or probable star \citep{Mongui}. For {\it Gaia}, the requirement summarizes as $|C^*| < 3 \, \sigma_{C*}$ where $C^*$ stands for the corrected excess flux factor in the $G_{RP}$ and $G_{BP}$ bands and $\sigma_{C*}$ is a power-law on the $G$ band with a chosen 3$\sigma$ level \citep{Riello}. For 2MASS, a good quality flag on each photometric band is required \citep{2MASS}. For UKIDSS, we impose $ErrBits < 256$, otherwise this indicated a bad value. We exclude as well photometry with either $J < 13.25$, $H < 12.75$ and $K < 12$, below which the photometry risks saturation \citep{UKIDSS}. 

Finally, we require at least one good photometric measurement in a blue band ($g$, $BP$ or $G$) and one in a near-infrared band, to ensure reliable SED fits, which reduces the size of our sample to 5,399,862 sources. 

To ensure the quality of the astrometry, we require $RUWE < 1.4$, leaving 5,211,036 sources. We go further by imposing $| \frac{\varpi}{q*\sigma_{\varpi}}  |> 2$ where $\sigma_{\varpi}$ is the  standard parallax uncertainty, and $q$ is a multiplicative factor that inflates the published standard uncertainty on parallaxes as it was found to be underestimated, notably for bright sources \citep{ElBaldry}. This leaves 2,145,918 sources. \footnote{\citet{Maiz2021} offer an alternative method for filtering {\it Gaia} data and correcting the astrometry that involves relaxing the constraint on RUWE and inflating the parallax uncertainties. We found that this approach typically increased the parallax errors by 10-20\%, but would have made our final sample 5\% larger (as the RUWE constraint is slightly relaxed). This change would therefore not have brought a significant change to our results.}

We showed in Section \ref{Known} that most of the stars in the Cygnus OB associations are located between 1 and 2.5 kpc. Therefore we removed all sources that fall outside this range of distances using the geometric values calculated by \citet{BailerEDR3}, trimming the sample to 896,946 sources.

Since we are interested primarily in OB stars, we limit our sample to sources with absolute magnitudes, $M_K < 1.07$, where $M_K = K - 5 \, \log(d) + 5$ with $d$ standing for the distance from \citet{BailerEDR3}, as the $K$ band is the least affected by extinction. This value corresponds to spectral types earlier than A0V \citep{Mamajek}. However, for sources that lack K-band photometry, we use $M_H < 1.10$ or $M_J < 1.07$, which offer equivalent criteria in the $J$ and $H$ bands. With these conditions, the sample is reduced to 47,498 sources. 

\begin{figure}
    \centering
    \includegraphics[width = \columnwidth]{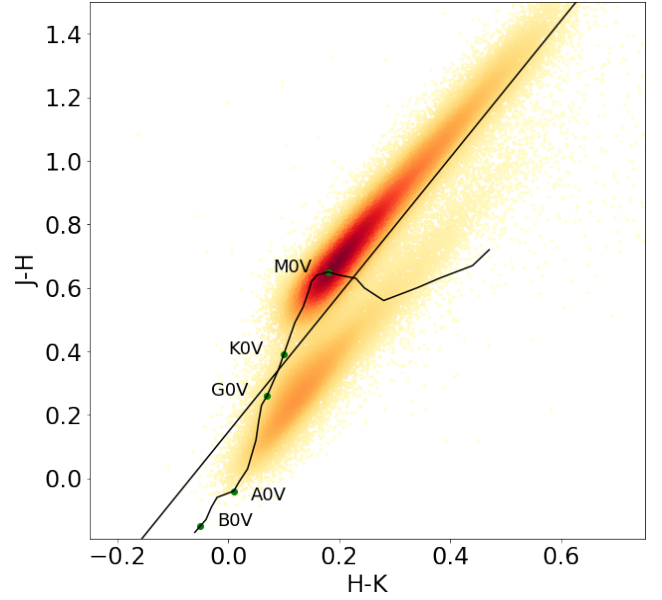}
    \caption{Near-IR colour-colour diagram displaying the source density from our photometric and astrometric sample. The black curve corresponds to the unreddened main-sequence from \citet{Straizys2009}. The black line, which is parallel to the near-IR reddening vector, is used to separate early-type stars of any reddening (below the line) from red giants (above the line). \label{IRSelec}}
\end{figure}

Finally we use the near-IR colour-colour diagram to remove giants by retaining only objects that have larger $H-K$ colours than the dashed line in Figure \ref{IRSelec} (projecting UKIDSS photometry into the 2MASS system using the equations from \citealt{Hodgkin}). This line separates early-type stars of any reddening from the majority of red giants. Sources that lacked sufficient photometry to be plotted in this diagram were retained regardless.

Following all cuts we were left with a sample of 20,498 objects that will be further classified using SED fitting. From the 152 sources listed by \citet{Blaha}, 87 passed all these tests.

\subsection{SED Fitting}
\label{SED}

In this section we describe our SED fitting process that is used to estimate the physical properties of our targets and subsequently select the OB stars we are interested in. We fit the observed SEDs and parallaxes to a series of model SEDs using a Markov Chain Monte Carlo (MCMC) process to explore the parameter space and constrain the posterior distribution. Here we describe the models used and the fitting process adopted.

\subsubsection{Model SEDs}
\label{modelSED}

Stellar SEDs can generally be parameterised by the combination of effective temperature, $T_{eff}$, luminosity, $L$, distance, $d$, and extinction, $A_V$. In some situations it can also be useful to parameterise the form of the reddening law used, for example using $R_V$. Main sequence models can be simplified by relying on a simple relationship (or model-predicted dependency) between $T_{eff}$ and $L$, though this can lead to uncertainties if the population to be fitted includes post-main sequence stars such as luminous red giants.

Instead of using $T_{eff}$ and $L$ to parameterise the unreddened SED, we utilise stellar evolutionary models that predict these parameters as a function of initial mass, $M$, and age. Furthermore, rather than using both distance and extinction as free parameters, we exploit the availability of 3D extinction maps, in this case the one produced by \citet{Bayestar}, to remove extinction as a free parameter and instead derive it from the distance\footnote{In early tests we found that using both $A_V$ and distance as free parameters lead to a systematic underestimation of $A_V$ compared to estimates from 3D extinction maps and that by requiring this dependency of one parameter on the other we overcame this.}. Finally we removed $R_V$ as a free parameter as studies in this area have shown that it does not vary significantly from the mean Galactic value of $R_V = 3.1$ \citep{Wright}.

We calculate our model SEDs using spectroscopy from two stellar spectral libraries, the Kurucz synthetic stellar library \citep{Coelho}, for models with $T_{eff}$ between 4000 and 20,000 K, and the Tubingen NLTE Model Atmosphere Package \citep{Werner1999,Rauch, Werner}, for models from 20,000 to 50,000 K. In each case, we chose models at $log(g) = 4$ (except for the model at $T_{eff} = 4000 K$ where we are forced to use $log(g) = 4.5$ instead), both using solar abundances. We will comment on the consequences of this choice in Section \ref{Compsib}. The model spectra are reddened using the \citet{Fitzpatrick} reddening model with $R_V$ set to 3.1, a value consistent with the region \citep{Wright}.

The reddened model spectra are then convolved  with the filter profiles introduced in Section \ref{Data}, to derive synthetic magnitudes \citep{MohrThesis}:
\begin{equation}
\label{Convolution}
m_x = -2.5 \, \log \bigg[\frac{\int \lambda \, F_{red} \, T_x \, d\lambda}{\int \lambda \, F_V \, T_x \, d\lambda}\bigg]
\end{equation}

\noindent where $T_x$ is the transmission curve of filter x and $F_V$ is a model Vega spectrum from \citet{Bohlin}.

To calibrate our model we calculated synthetic colours for main-sequence stars and compared them to the empirical main sequence from \citet{Mamajek} for {\it Gaia} and 2MASS and \citet{Verbeek} for IGAPS. We found generally excellent agreement: the root mean squares differences for the colours ($g-r$, $BP-RP$, $G-RP$, $J-H$, $H-K$) were equal to (0.025, 0.012, 0.037, 0.005). The only band for which a discrepancy was noted was the $H$ band, for which we needed to add 0.0375 to all magnitudes to bring them in line with empirical values. We attribute this to either uncertainties in the modelled OH lines in stellar spectra or the H-band filter transmission function.

\subsubsection{Fitting process}
\label{fitprocess}

The list of model parameters we are seeking to estimate are:
\begin{equation}
\theta = [\log(Mass), Fr(Age), d, \ln(f)]
\end{equation}

\noindent where $Fr(Age)$ stands for the fractional age and $\ln(f)$ represents the logarithm of an additional uncertainty that needs to be considered to achieve agreement between the observations and the model. Adding such a parameter helps $\chi^2$ to converge to 1 for the computation of the likelihood \citep{Emcee}. The new errors will therefore be expressed as (e.g. \citealt{Casey}):
\begin{equation}
\label{lnf}
s^2 = \sigma^2 + f^2 \, Model^2
\end{equation}

\noindent where $Model$ is the combination of the model SED and the model parallax. 

We  define the following priors for these parameters:
\begin{equation}
\ln (P(\theta)) =
\begin{cases}
\log (\frac{1}{2 \, L^3} \, d^2 \, \exp{(\frac{-d}{L}})) & \text{if }
\begin{cases}
-1.0 \leq \log(Mass) \leq 2.0  \\
0.0 \leq Fr(Age)  \leq 1.0 \\
0.0 \leq d \leq 5000.0 \,  pc \\
-10.0 \leq \ln(f) \leq 1.0 \\
\end{cases} \\ 
- \infty  & \text{otherwise}
\end{cases}
\end{equation}

\noindent The prior on distance originates from \citet{Bailer2015}, and is designed to achieve an exponentially decreasing volume density of stars, using a scale length of L = 1.35 kpc. 

The parameters of the model SED are constrained by comparing them to the observed SED using Bayesian inference and a maximum likelihood test. Through this method, with a set of empirical data and model parameters, we can compute the posterior probability, labelled $P(\theta|d)$, which is the probability of a set of model parameters, given the observations, by applying Bayes' theorem:

\begin{equation}
P(\theta|d) \propto P(d|\theta) \, P(\theta)
\end{equation}

\noindent where $P(\theta)$ are the priors and  $P(d|\theta)$, named the likelihood, corresponds to the probability that the data are measured given the model parameters.

The observational data, consisting of the photometry and astrometry, and their uncertainties, can be written as:

\begin{equation}
Obs = [g, BP, r, G, i, RP, J_{2M}, J_U, H_U,H_{2M}, K_{2M}, K_U, \varpi]
\end{equation}
\begin{multline}
\label{sigma}
\sigma = \bigg[\sigma_g, \sigma_{BP}, \sigma_r, \sigma_G, \sigma_i, \sigma_{RP}, \\ \sigma_{J_{2M}},\sigma_{J_U}, \sigma_{H_U}, \sigma_{H_{2M}}, \sigma_{K_{2M}},\sigma_{K_U}, \sigma_{\varpi}\bigg]
\end{multline}

\noindent where the $2M$ subscript stands for 2MASS and $U$ for UKIDSS. The photometric uncertainties are the combination of their standard measurement uncertainty, and a systematic error. The value of the latter is taken as 0.03 mag for $g$, $r$ and $i$ \citep{Drew2014, BarentsenArt},  0.01 mag for $G$, $G_{BP}$, $G_{RP}$ \citep{Riello}, 0.03 mag for $J_{2M}$,0.02 for $H_{2M}$ and $K_{2M}$ \citep{Skrutsie}, and 0.03 mag for $J_U$, $H_U$, $K_U$ \citep{Hodgkin}.

The best fit model will thus be that which minimizes the negative logarithm of the likelihood: 

\begin{equation}
\ln (P(Obs | \theta)) = - \frac{1}{2} \, \sum_{i}^{n} \frac{(Obs(i) - Model(i))^2}{s_i^2} + \ln (s_i)
\end{equation}

\begin{figure*}
\includegraphics[scale = 0.10]{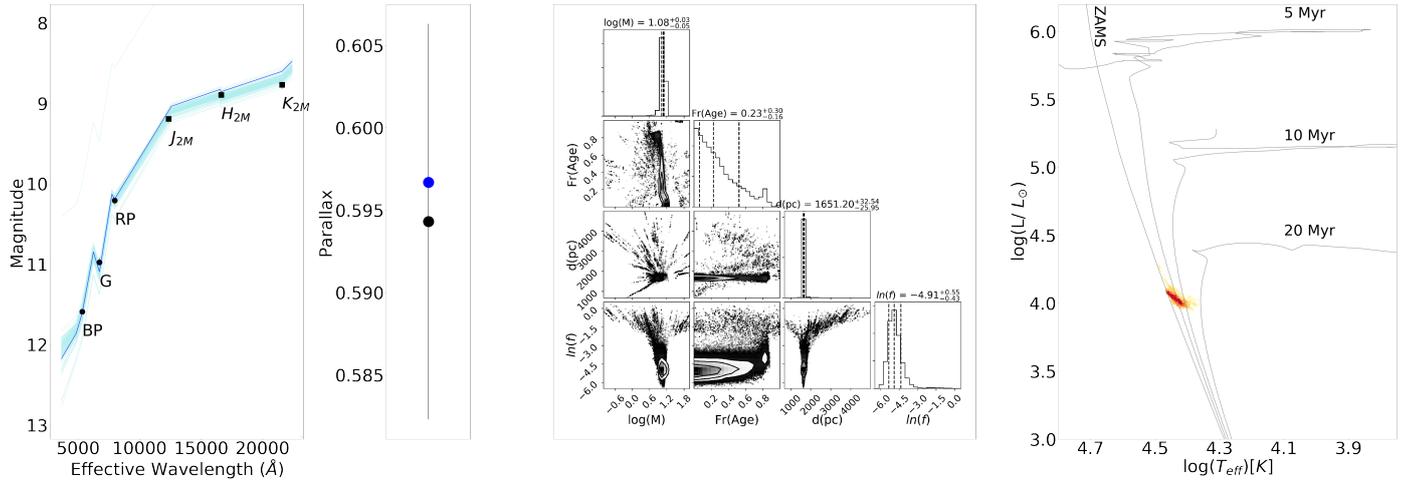}
\caption{Results of the fit for J203227.75+41285.2 (Schulte 21), fitted with $log(T_{eff}) = 4.44 \pm 0.02$, compared with a spectroscopic $log(T_{eff}) = 4.45^{+0.03}_{-0.05}$ \citep{Kiminki}. Left: SED fit with the data shown in black and inflated error bars.  The blue line shows the best fit model whilst the cyan lines correspond to 100 randomly-selected iterations from the posterior distribution. Centre-left: Fitted parallax in blue vs \textit{Gaia} parallax in black. Centre-right: Triangle plot showing the posterior distributions of the fitted parameters. Right:  Posterior distributions of $log(T_{eff})$ and $log(L/L_{\odot})$ shown as a heat map in an HR diagram  with a zero-age main-sequence and isochrones from \citet{Ekstrom}. \label{Corner}}
\end{figure*}

To  explore the posterior distribution and identify the best-fitting parameters, we use the \texttt{emcee} MCMC package in Python \citep{Emcee}. For the MCMC simulation we use 1000 walkers, which undergo 100 burn-in iterations and then 1000 iterations to fully sample the posterior distribution. The fitted parameters are taken as the median of the resulting posterior distribution, while the 16th and 84th percentile values provide the lower and upper 1$\sigma$ bounds. We also extracted fits and uncertainties for the intermediate parameters, $log(T_{eff})$ and $log(Lum)$. The median of the posterior distribution was preferred over the best-fit value for all of these quantities as it showed better agreement for objects with spectroscopic effective temperatures (see Section \ref{Compsib}). An example of the posterior distributions, SED fit and posterior distributions of effective temperature and luminosity in the HR diagram are shown in Fig. \ref{Corner}.

\subsubsection{General results}
\label{results}

\begin{figure*}
    \centering
    \includegraphics[scale = 0.08]{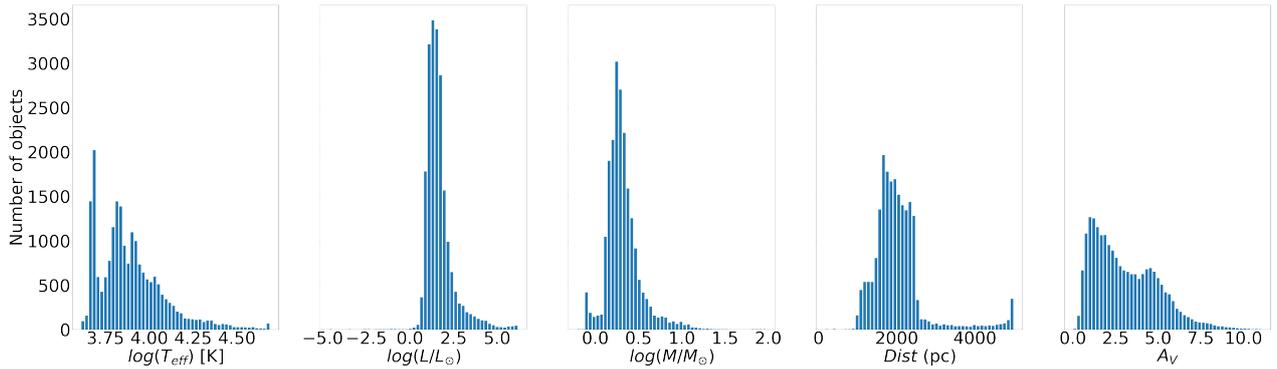}
    \caption{Distribution of the fitted physical parameters derived from the SED fitting process for the 20,498 sources.   \label{Distrib}}
\end{figure*}

SED fits were performed for 20,498 objects and the distribution of the physical parameters is displayed in Fig. \ref{Distrib}. 

The distribution of $T_{eff}$ shows that the sample is clearly dominated by cool red giants, despite our previous cuts, but we also find that 4680 objects have $log(T_{eff}) > 4$ (OB stars, 22.8\%), and 818 have $log(T_{eff}) > 4.3$ (O stars, 4.0\%). The distributions of $log(L/L_{\odot})$ and $log(M/M_{\odot})$ reveal, respectively, peaks between 10 and 100 $L_{\odot}$ and between 1 and 3 $M_{\odot}$, another strong indicator of the dominance of cool giants. 

As expected from our distance cuts, most of our fitted distances fall between 1 and 2.5 kpc, with a peak between 1.5 and 2 kpc. Sources fitted close to the upper prior on distance (5 kpc) are often indicators of poor fits.  The reddening for our fitted objects, as derived from the reddening maps of \citet{Bayestar} shows a broad spread from $A_V$ = 0 to 6.

\subsubsection{Comparison with known results}
\label{Compsib}

We cross-matched our sample with the \textit{Simbad} database with a matching radius of 1 arcsec and found 2672 matches. From this list, we focus only on those with a clearly identified spectral type, a reference, and with a quality measurement of 'A', 'B' or 'C' ('D' and 'E' are considered too low quality). 

We determine the effective temperatures from the spectral type using the tabulations from \citet{Martins} for the O-type stars (observed scale), from \citet{Trundle} for early B-type stars, from \citet{Humphreys1984} for late B-type stars of luminosity classes 'I' or 'III' and from \citet{Mamajek} for the others, interpolating between these tabulations where they overlapped. We required at least a spectral type and a spectral subclass, used the spectral type of the primary star for binaries, assumed a luminosity class of 'V' when not specified, interpolated for luminosity classes 'II' and 'IV', and chose error bars of one spectral subclass. This process reduces the number of stars with spectroscopic effective temperatures to 205 objects, including 44 from \citet{Blaha} and 60 from \citet{Berlanas2}.

\begin{figure}
    \centering
    \includegraphics[scale = 0.4]{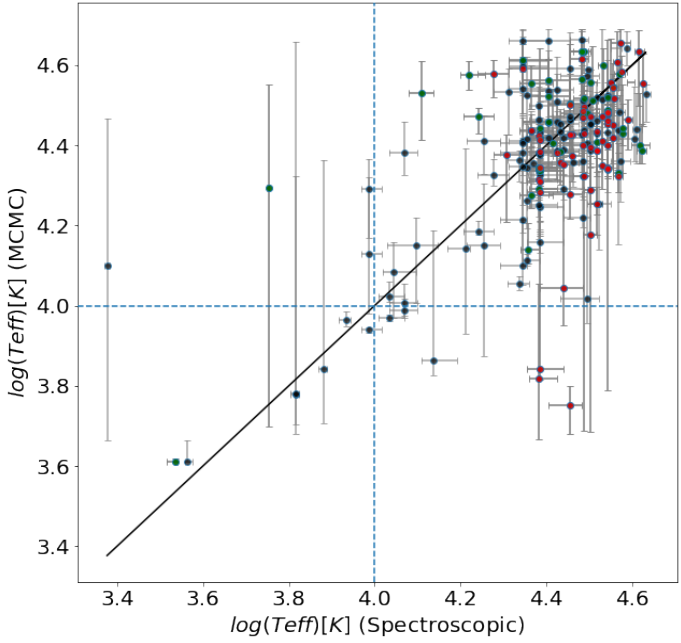}
    \caption{Comparison between the literature spectroscopic temperature with that from our SED fits, for the 205 objects with reliable spectral types. Blue dashed lines correspond to the 'limit' at $log(T_{eff}) = 4$ (dividing OB stars from cooler stars). Green sources are from \citet{Blaha} while red ones are from \citet{Berlanas2}. \label{Compteff}}
\end{figure}

Fig. \ref{Compteff} shows a comparison between the spectroscopic effective temperatures and those from our SED fits, and reveals that very few cool stars are identified as hot stars by our SED fitting process. We define our recovery rate as $RR = \frac{TP}{TP+FN}$ where $TP$ stands for the number of true positives and $FN$ corresponds to the number of false negatives. Our recovery rate is 97\% (96\%, 92\% and 86\%) for stars with SED-fitted $log(T_{eff})$ > 4.0 (4.1, 4.2, 4.3), and our contamination rate ($\frac{FP}{TP+FP}$, where $FP$ is the number of false positives) is between 3 and 14\%, depending on the temperature threshold used. 

Exploring our fits in more detail, we found that sources with a large extinction ($A_V > 4$ mag) typically have higher $ln(f)$ values (> -5) and that these corresponded to stars with poorer fitted temperatures (when compared to literature values) and larger uncertainties on the fitted quantities. As an example of this, if we focus our comparison on the 44 objects from \citet{Blaha} with literature spectral types (typical extinction of $A_V$ = 1-2 mag), we find a recovery rate of 92--100\%, while if we limit our comparison to the 60 objects in Cyg OB2 (typical extinction of 4-5 mag) listed by \citet{Berlanas2} our recovery rate drops to 85--95\%, slightly inferior. Fortunately, our focus is not on the high-extinction region of Cyg OB2, but on the other OB associations that exhibit a lower extinction, meaning that our recovery rate for OB stars is approximately 92-100 \%.

Finally, we explored the impact of using a single value of $log(g)$ in our fitted stellar spectral models. This was necessary because of the difficulty finding models covering the full range of $T_{eff}$ and $log(g)$, and with a sufficient wavelength coverage. We estimated the bias that using a fixed value of $log(g)$ introduced by comparing the effective temperatures and luminosities derived from our SED fits with those derived spectroscopically, for the 199 sources with spectroscopic surface gravities. We divided this sample into stars with low ($log(g) < 3.5$) and high ($log(g) > 3.5$) surface gravity. We found that stars with a low surface gravity had their SED-fitted effective temperatures over-estimated by $\sim$0.1 dex, while their luminosities were under-estimated by $\sim$0.01 dex. Both of these effects will cause evolved stars with low surface gravities to be fitted closed to the ZAMS than they should be, under-estimating their ages.

\section{Identifying new OB associations}
\label{newassoc}

In this section we exploit our sample of OB stars to search for OB associations in the Cygnus region. For this purpose, we will use a new method, the results of which will be compared with existing clustering algorithms.

\subsection{Broad kinematics distribution of OB stars}
\begin{figure}
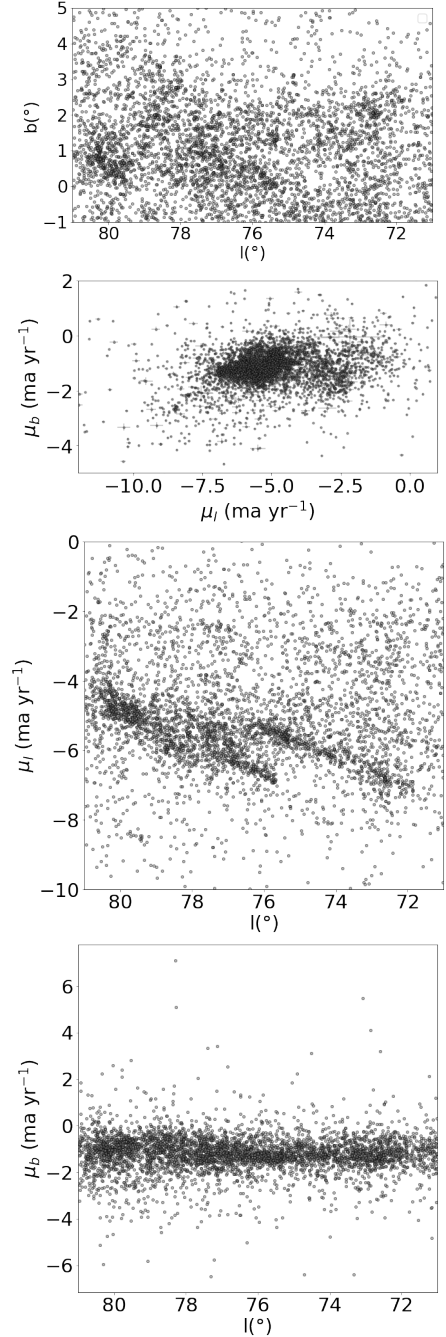

    \centering
    \includegraphics[scale = 0.24]{lbGen.pdf}
    \includegraphics[scale = 0.24]{PmGen.pdf}
    \includegraphics[scale = 0.24]{lvspmlGen.pdf}
    \includegraphics[scale = 0.24]{lvspmbGen.pdf}
    \caption{Spatial and proper motion distributions of the 4680 objects fitted with $log(T_{eff}) > 4$. \label{Broadkin}}
\end{figure}

Fig. \ref{Broadkin} shows the broad spatial and kinematic distribution of our SED-fitted OB stars ($log(T_{eff}) > 4$). The spatial distribution (top panel) shows a broad density gradient as a function of Galactic latitude, as expected, with multiple over-densities that might represent OB associations (including one in the position of Cyg OB2 at $l \sim 80^{\circ}$,  $b \sim 1^{\circ}$). The proper motion distribution (second panel) shows that the bulk of the stars have proper motions between $\mu_l = -8$ and $-2$ mas yr$^{-1}$ and around $\mu_b \sim -1$ mas yr$^{-1}$. The graph displaying $l$ vs $\mu_l$ (centre panel) reveals two diagonal structures, above and below $l = 76^{\circ}$, an interesting saw-tooth pattern indicating a correlation between position and velocity that is usually indicative of some form of expansion (we will return to this feature on Section \ref{analysis}). No such feature was observed in $\mu_b$ (see bottom panel of Fig. \ref{Broadkin}).

The expected motion of stars in this sightline can be estimated using a simple Galactic rotation model. \citet{Eilers} determined a Galactic rotation curve of $v = 229 \, km \, s^{-1} \, - 1.7 \, km \, s^{-1} \, kpc^{-1} (R - R_{\odot})$. Along our line of sight these stars have a Galactocentric radius of $R \sim  7.7$ kpc, if we assume d = 1.8 kpc and $l = 72^{\circ}$, at an angle intercepting the line of sight of 4.5 degrees we have a rotation speed of 229.6 km s$^{-1}$. This equates to a motion of about -18 km s$^{-1}$ in the $l$ direction. With the local Solar motion of U = 11.1 km s$^{-1}$ and V = 12.24 km s$^{-1}$ \citep{Schon}, the Sun requires a correction of about -6.73 km s$^{-1}$ in that direction. This therefore results in an apparent motion of -24.73 km s$^{-1}$, or a proper motion of about -2.74 mas yr$^{-1}$ assuming a distance of d = 1.8 kpc. The majority of stars in this direction have a proper motion in $l$ more negative than this, suggesting that the Cygnus region is moving towards lower $l$ values (towards the inner Galaxy) relative to other stars orbiting within the Milky Way.

Stars in this sightline appear to have a typical proper motion in Galactic latitude of -1 mas yr$^{-1}$. If the motion of the Sun is 7 $\pm$ 0.5 km s$^{-1}$ \citep{Bland}. in the positive $b$ direction, then stars in the Galactic disk will have negative proper motions of a similar magnitude. At a distance of 1.8 kpc this would equate to $\sim$ -1 mas/yr, consistent with what is observed in this sightline.

\subsection{Identifying kinematic groups}
\label{ktest}

For the purpose of identifying OB associations we limit our sample of stars to those  with $log(T_{eff}) > 4.2$ (approximately equivalent to spectral type B5 or earlier), which gives us 1349 stars. This limits us to a sample of more massive and therefore younger stars with which to identify young groups.

To identify OB associations or groups of kinematically-coherent stars within the wider distribution of OB stars, we seek to find groups of stars relatively close together on the sky whose kinematics are more similar to each other than to the wider distribution of OB stars. To start we want to visualise how the relative kinematics of stars changes over our area of study. To do this we define a grid with a cell size of 0.1$^{\circ}$ and at each point in this grid we select the ten nearest stars from the sample of 1349 stars. We then perform a Kolmogorov-Smirnov (KS) test comparing the proper motion distribution of these ten stars with the proper motion distribution of all 1349 stars. We do this for both the proper motion in $l$ and in $b$, obtaining a P-value for each, and then multiply them together. The result is the probability that the kinematics of stars in that area are consistent with that of the wider population. 

\begin{figure*}
    \centering
    \includegraphics[scale = 0.4]{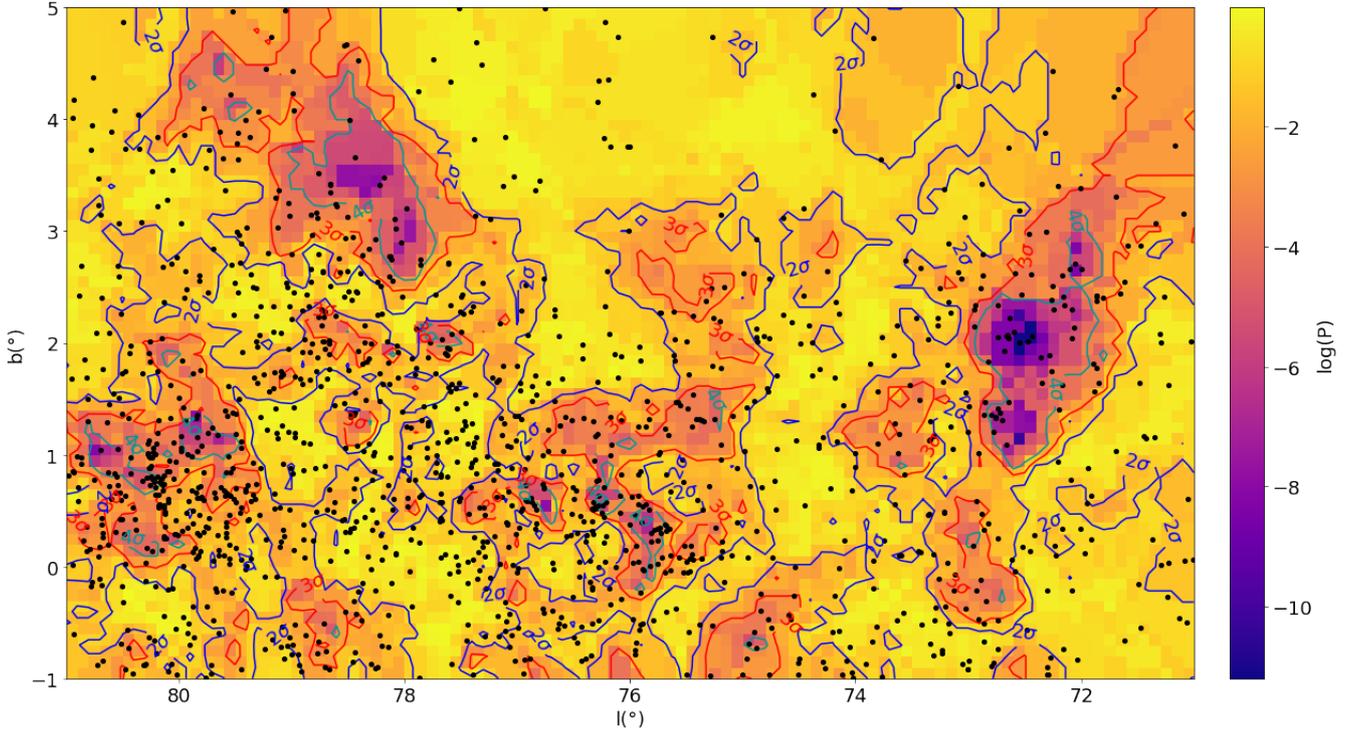}
    \caption{The Cygnus region, colour-coded according to probability, log(P), that the kinematics of stars in that vicinity are consistent with that of the wider population. The 2$\sigma$, 3$\sigma$, 4$\sigma$ contours represent respectively levels at log(P) = -1.3,-2.6 and -4.2. \label{Pvalue}}
\end{figure*}

Figure \ref{Pvalue} shows the distribution of this probability, over our area of study, effectively highlighting regions that are kinematically distinct compared to the wider area. Contours are shown at 2, 3, and 4$\sigma$ significance. Immediately apparent is that there are a number of areas with very significantly distinct kinematics ($\mathrm{log}(P) < -6$), one of which is located in the direction of Cyg OB2 ($l = 80^{\circ}$, $b = +1^{\circ}$), giving an early verification of our method.

\begin{figure}
    \centering
    \includegraphics[width = \columnwidth] {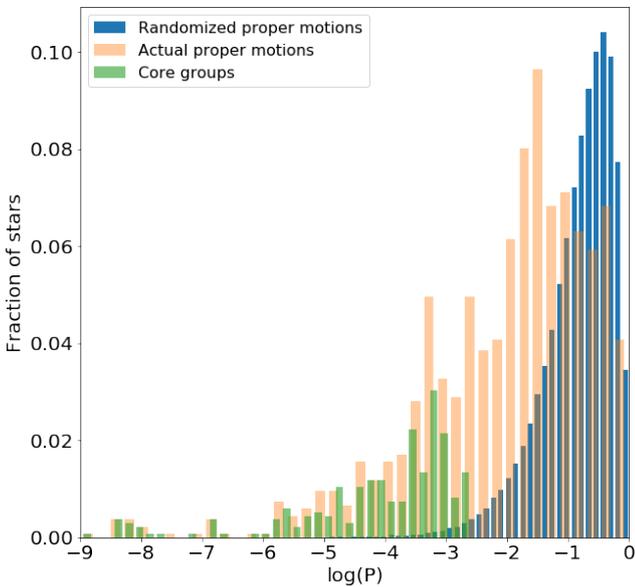}
    \caption{Log(P) values across our survey area deduced from the randomized proper motions, from the actual proper motions, and for the core groups (before removal of outliers). \label{HistlogP}}
\end{figure}

To test our method and the significance of our results we perform a simple experiment, repeating our analysis 100 times, but in each iteration randomizing the proper motions of all the stars and calculating the log(P) value at each position. The goal of this process is to check whether the probabilities derived from the actual proper motions are consistent with a random distribution of kinematics. Fig. \ref{HistlogP} compares the observed distribution of probabilities with that derived from the randomised velocities and shows the former reaches considerably smaller values of log(P). This suggests that the majority of the kinematic structures observed in Fig. \ref{Pvalue} are likely to be real.

To identify stars in a given group, we choose the 3-sigma contour level and define all stars falling within such a contour as being tentatively part of that group. Given the number of contours, we required each group to contain a minimum of 10 stars, which led to the identification of five groups of stars with similar kinematics. This method of identifying groups is likely to include some contaminants that are projected against real groups. To exclude any such contaminants we then calculated the median and 16th and 84th percentile values of  $\mu_l$, $\mu_b$ and parallax, and excluded sources that were more than 3$\sigma$ from the median value in each dimension, where $\sigma$ is the true dispersion calculated using the outlier-resistant method from \citet{Ivezic}. This process reduced the membership of each group by approximately 20--30\%.

From this we find the following groups: 
\begin{enumerate}
    \item Group A is located in the upper-right part of Fig. \ref{Pvalue}, with $l = 71.5-73^{\circ}$ and $b = 0.5-3.0^{\circ}$.
    \item Group B spans the upper-left part of Fig. \ref{Pvalue}, in the region $l > 77-80^{\circ}$ and $b = 2.5-5.0^{\circ}$.
    \item  We initially defined a third group in the region of l from 75 to 77.5$^{\circ}$ and b from -0.5 to 2$^{\circ}$, but further investigation (see later) suggested this constituted two groups with distinct $\mu_l$ distributions and so this was separated into two groups, C and D, with $\mu_l$ less than and greater than $\mu_l = -6$ mas/yr, respectively.
    \item The group in the lower left part of Fig. \ref{Pvalue} corresponds to Cyg OB2 and its surrounding. It forms group E.
    \end{enumerate}
    
Our initial selection of group members is based on stars with $log(T_{eff}) > 4.2$, chosen as the balance between a large enough sample and stars that are sufficiently hot to be very massive and young. With our groups defined we can now expand their membership by considering the inclusion of slightly cooler stars. We add stars that fulfill either $log(T_{eff}) > 4$ (and $< 4.2$) or $log(\frac{L}{L_{\odot}}) > 2.5$, hence we catch all possible OB stars, and massive post-main sequence stars that have evolved to cooler temperatures. We add to each group any star that fulfils these criteria, falls within the same 3$\sigma$ contours, and is within 2 standard deviations of the median value of parallax and proper motion (with the median and standard deviations defined from the current group membership). This increased the membership of the existing groups to a total of 123, 93, 93, 86 and 163 stars. 

\begin{figure*}
    \centering
    \includegraphics[scale = 0.3] {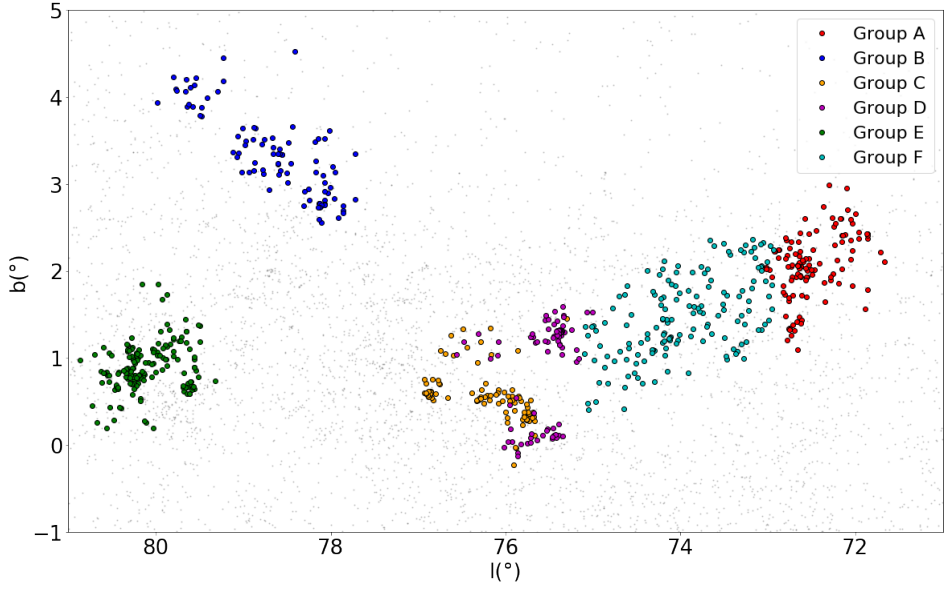}
    \caption{Spatial distribution of all the identified new OB groups (coloured symbols), plotted against the field OB stars (black dots). \label{SpaceDistribExtended}}
\end{figure*}

\begin{figure*}
    \centering
    \includegraphics[scale = 0.4] {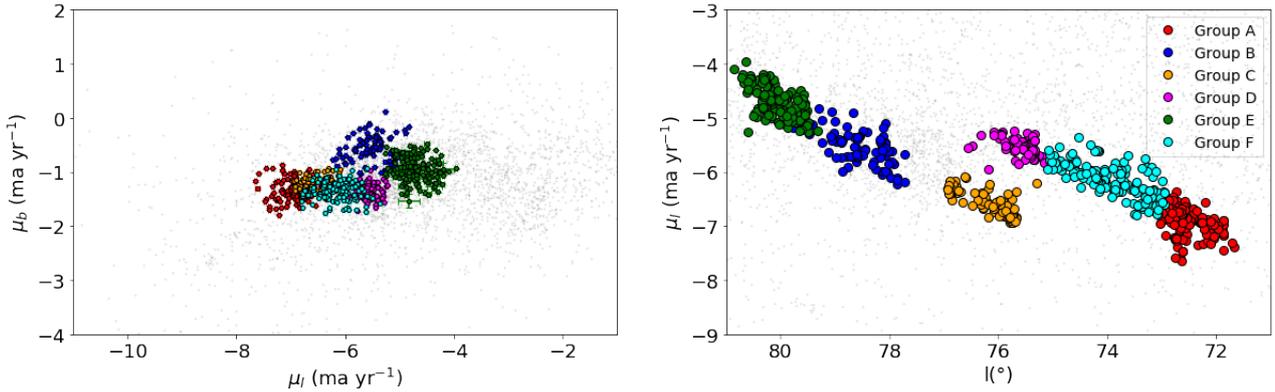}
    \caption{Left: proper motion distribution of all 6 new OB groups (coloured symbols), plotted against the field OB stars (black dots). Right: proper motion in $l$ plotted against $l$ for the same objects. The uncertainties in $\mu_l$ are comparable to, or smaller than, the symbol size, and thus aren't shown. \label{PmDistribExtended}}
\end{figure*}

Finally, in later analysis of the kinematics of these stars we became aware of  a 'gap' between groups A and D in their distributions in $l$, $b$, and $\mu_l$. We identified the stars that fell between groups A and D in both $l$, but also in the $l$ vs $\mu_l$ diagram (see Fig. \ref{PmDistribExtended}), trimmed the sample and then added in slightly cooler stars, all as described above (this time also discarding the outliers in $l$ and $b$). This resulted in a group of 155 stars. It should be noted that the group has been selected manually, contrary to the other groups. Fig. \ref{SpaceDistribExtended} shows the spatial distribution of all the final groups, while Fig. \ref{PmDistribExtended} shows the proper motion distribution (left panel) and $l$ vs $\mu_l$ (right panel) distribution of the groups. 

\subsection{Comparison with historical associations}
\label{HistCompSection}

How do our new groups compare to the historical associations discussed in Section \ref{Known}? We compared the membership of our groups with that of the historical associations, our results confirming what we found in Section \ref{Known}: while Cyg OB2 stands out as a real group (corresponding roughly to our group E) and Cyg OB3 has some overlap with group A, the same is not true for the other groups, which show very little overlap with the historical OB associations (typically 5-20\% of the historical association members are found in the nearest new groups we identify). Comparison plots between the historical associations and our new groups are shown in Fig. \ref{CompHist}.

\subsection{Comparison with the results from DBSCAN}
\label{DBSCANComp}

To test our group-finding method and compare the results to those from modern clustering algorithms, we again selected the 1349 stars of our sample with $log(T_{eff}) > 4.2$ and used DBSCAN \citep{DBS} to identify groupings, choosing the number of nearby stars to be ten for similarity with our own method and using a maximum normalised distance between stars (in both spatial and proper motion coordinates) to be considered part of the same group of 0.05.

DBSCAN was able to identify six groups, respectively containing 158, 167, 23, 52, 12 and 32 stars. Fig. \ref{CompDBSCAN} shows how these groups compare to our groups, some of which match up well, while others don't. For example, our group A agrees well with DBSCAN group 6 and our group E matches with DBSCAN group 2, though in both cases our groups are more extended than those identified by DBSCAN. In other cases the agreement is less good, our groups C and D overlap with DBSCAN groups 1 and 3 in a complex manner. The most obvious distinction lies between group 5 from DBSCAN and our group B. DBSCAN was unable to pick out the bulk of group B. This is because, as outlined above, one of the main parameters of DBSCAN is the relative distance between the objects used to form clusters. Consequently, since the same value was used throughout the entire region, it was impossible to detect this lower-density group, while still detecting the higher-density groups found elsewhere. Our method, on the other hand, only relies on the number of neighbours stars, and therefore isn't biased by density in this regard.

Sometimes the opposite trend is observed, i.e. some stars have been identified as parts of the groups in DBSCAN but not with our method. This stems from our selection: we chose sources within 3-sigma contours, rejecting sources that fell outside of these contours and which may represent members in the periphery of a group.

\section{Analysis of the new OB associations}
\label{analysis}

In this section we analyse the physical and kinematic properties of our new groups. To do so, we calculate several of their properties and produce HR diagrams of their members in order to obtain broad estimates of their age. We then give a few words about their dynamics. Table \ref{TabParamGroups} summarises some of the physical and kinematic properties of each group, which were calculated as follows.

\begin{table*}
	\centering
	\caption{Properties of our new OB associations. The first column indicates the parameter, where the subscript 'm' indicates the median value and '$\sigma$' the dispersion. \label{TabParamGroups}}
	\renewcommand{\arraystretch}{1.3} 
	\begin{tabular}{lccccccccr} 
		\hline
		Parameters & Units & Group A & Group B & Group C & Group D & Group E & Group F\ \\
		\hline
        ${RA(ICRS)}_m$ & deg & 301.45 & 304.37 & 305.47 & 304.34 & 308.08 & 302.95 \\
        ${DE(ICRS)}_m$ & deg & 35.68 & 41.43 & 37.87 & 37.64 & 41.30 & 36.58 \\
        $l_m$  & deg & 72.61  &  78.58 & 76.11 & 75.44 & 80.19 & 74.04 \\
        $b_m$  & deg & 2.06 & 3.31 & 0.54 & 1.19 & 0.85 & 1.44 \\
        $d_m$ & pc & 1894.5 & 1726.3 & 1713.1 & 2000.1 & 1674.0 & 1985.2 \\
        $\mu_{l_m}$  & mas yr$^{-1}$& -6.90 & -5.47 & -6.57 & -5.55  & -4.72 & -6.11 \\
        $\sigma_{\mu_l}$  & mas yr$^{-1}$& 0.24 & 0.34  & 0.27 & 0.16 & 0.27 & 0.30 \\
        $\mu_{b_m}$ & mas yr$^{-1}$ & -1.35 & -0.59 & -1.19 & -1.34 & -0.96 & -1.33 \\
        $\sigma_{\mu_b}$ & mas yr$^{-1}$ & 0.17 & 0.27 & 0.12 & 0.14 & 0.25 & 0.1 \\
        $\sigma_{v_l}$ & km s$^{-1}$ & $2.20^{+0.55}_{-0.47}$ & $2.72^{+0.47}_{-0.37}$ &  $2.24^{+0.46}_{-0.35}$ & $1.53^{+0.45}_{-0.41}$ &  $2.15^{+0.66}_{-0.45}$ &  $2.80^{+0.58}_{-0.48}$ \\
        $\sigma_{v_b}$ & km s$^{-1}$ & $1.55^{+0.32}_{-0.24}$ & $2.18^{+0.41}_{-0.33}$ &  $0.96^{+0.22}_{-0.18}$ &  $1.32^{+0.30}_{-0.26}$ &  $1.96^{+0.52}_{-0.29}$ &  $1.50^{+0.30}_{-0.25}$ \\
        Observed number of B stars & & $112^{+3}_{-3}$ & $82^{+3}_{-2}$ & $76^{+2}_{-3}$ & $75^{+2}_{-3}$ &  $110^{+4}_{-3}$ & $139^{+3}_{-3}$ \\
        Observed number of O stars & & $3^{+1}_{-1}$ & $4^{+1}_{-1}$ & $10^{+1}_{-2}$ & $4^{+2}_{-1}$ & $35^{+2}_{-3}$ & $5^{+2}_{-1}$ \\
        Corrected number of O stars & & $13^{+4}_{-3}$ & $11^{+4}_{-3}$ & $12^{+4}_{-3}$ & $11^{+4}_{-3}$ & $23^{+6}_{-5}$ & $17^{+4}_{-5}$ \\
        Estimated total stellar mass & $M_{\odot}$ & $2344^{+275}_{-252}$ & $1978^{+245}_{-224}$ & $2165^{+245}_{-224}$ & $1569^{+215}_{-192}$  & $4198^{+354}_{-319}$ & $2955^{+349}_{-269}$  \\
        Velocity gradient (l) &  mas yr$^{-1}$ deg$^{-1}$ & $0.24 \pm 0.07$ & $0.39 \pm 0.04$ & $0.38 \pm 0.04$ & $0.20 \pm 0.05$ & $0.43 \pm 0.06$ & $0.35 \pm 0.03$ \\
        Velocity gradient (l) &  km s$^{-1}$ pc$^{-1}$ & $0.07 \pm 0.02$ & $0.11 \pm 0.01$ & $0.10 \pm 0.01$ & $0.05 \pm 0.01$ & $0.12 \pm 0.01$ & $0.09 \pm 0.01$ \\
        Velocity gradient (b) &  mas yr$^{-1}$ deg$^{-1}$ & $0.34 \pm 0.04$ & $-0.03 \pm 0.04$ & $0.10 \pm 0.04$ & $0.07 \pm 0.03$ & $0.16 \pm 0.05$ & $0.06 \pm 0.03$ \\
        Velocity gradient (b) & km s$^{-1}$ pc$^{-1}$ & $0.09 \pm 0.01$ & $-0.01 \pm 0.01$ & $0.03 \pm 0.01$ & $0.02 \pm 0.01$ & $0.04 \pm 0.01$ & $0.02 \pm 0.01$ \\
        Expansion age (l) & Myr & $13.98 \pm 3.99$ &  $8.89 \pm 0.81$ & $7.93 \pm 0.80$ & $19.57 \pm 3.91$ & $8.37 \pm 0.70$ & $10.87 \pm 1.21$ \\
        Expansion age (b) & Myr & $10.87 \pm 1.21$ & - & $32.62 \pm 10.87$ &
        - & $24.46 \pm 6.11$ & - \\
		\hline
	\end{tabular}
\end{table*}

\subsection{Physical properties of the individual groups}
\label{individual}

To estimate the observed number of O and B stars in each group, we defined  B-type stars as those with $log(T_{eff}) > 4$ and $log(T_{eff}) < 4.47$ and  O-type stars as those with $log(T_{eff}) > 4.47$. Where a spectral type is available from the literature we use the spectroscopic $T_{eff}$, otherwise we use the effective temperature derived from our SED fits. The uncertainties on the number of O and B stars was estimated using a Monte Carlo experiment, varying the temperature of each star according to its uncertainties.

We then attempt to calculate the total mass of each group. We used the SED-fitted masses for consistency. To determine the complete mass range, we inspected the mass function, chose the turn-over point of the mass function as the lower limit and the post-main sequence turn-off mass at 15 Myr for the upper limit \citep{Ekstrom}. Based on this we estimate our samples to be relatively complete in the mass range of 3.2 -- 13.8 $M_{\odot}$.

\begin{figure}
    \centering
    \includegraphics[scale = 0.4]{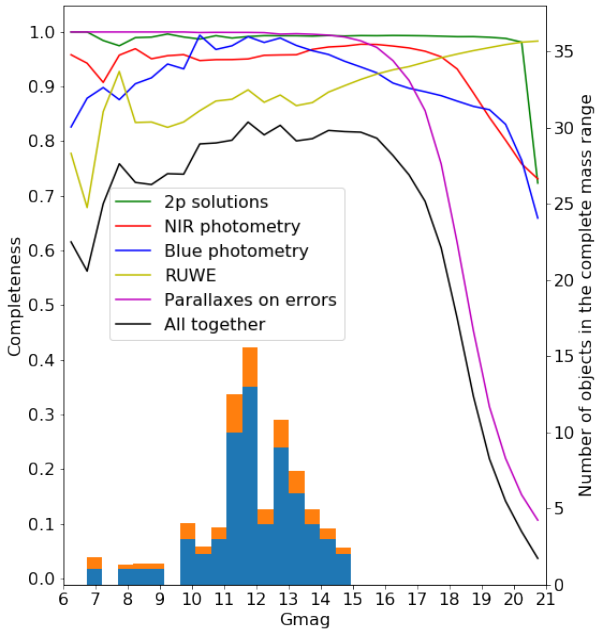}
    \caption{Completeness as a function of $G$ magnitude for the various data filtering steps in Section \ref{Data}. The black curve corresponds to the convolution of all effects combined together. The histogram show the number of stars in group A within the complete mass range both before (blue) and after (orange) applying the incompleteness corrections. \label{FractionPass}}
\end{figure}

The filtering of the data we used (Section \ref{Data}) means that our samples of members will be slightly incomplete. We estimate our incompleteness by calculating the fraction of stars, as a function of magnitude, that were discarded at each step in Section \ref{Data}. The completeness level varies for each step, and as a function of magnitude (shown in Fig. \ref{FractionPass}), with a cumulative completeness of 70--80\% in our magnitude range of interest. Fig. \ref{FractionPass} shows these incompleteness curves and an example of how the membership of one of our groups increases once this incompleteness is accounted for.

The last source of incompleteness stems from the false positive and negative rates from our SED fitting process. In Section \ref{Compsib}, we quantified the recovery rates at different temperature thresholds by comparing our SED-fitted temperatures with literature spectroscopic temperatures (see Fig. \ref{Compteff}). We repeat this here using as a threshold the minimum mass of our complete mass range ($log(T_{eff} = 4.11$). For sources with $A_V < 4$ (appropriate for the low-extinction associations) our false negative rate is 5.0\% and our false positive (contamination) rate is 5.9\%, while for sources with $A_V > 4$ (appropriate for Cyg OB2), these quantities are respectively equal to 3.5\% and 0.0\%.  We then estimate the 'true' number of stars, $T$, from:
\begin{equation}
\centering
T = \frac{N}{1-F} - N \, P
\end{equation}
\noindent where $N$ is the corrected number of stars within the complete mass range, $F$ is the false negative rate and $P$ the false positive rate. 

To estimate the total mass of each group, we apply both of these corrections. Subsequently, to estimate the total mass of each group we perform a Monte Carlo simulation using the mass functions from \citet{IMFMasch} to sample stellar masses, counting the number of stars in our representative mass range as well as the total number (and mass) of stars. When our simulation reaches the total number of observed stars in our representative mass range we halt the simulation and take the total mass of the group of stars. We repeat this 10,000 times, varying the number of stars observed in the representative mass range (according to their uncertainties), to estimate the uncertainties on the total mass. We use this same method to deduce the incompleteness corrected number of O-type stars, also listed in Table \ref{TabParamGroups}.

\subsection{HR Diagrams and ages of group members}
\label{HRs}

\begin{figure*}
    \centering
    \includegraphics[scale = 0.18]{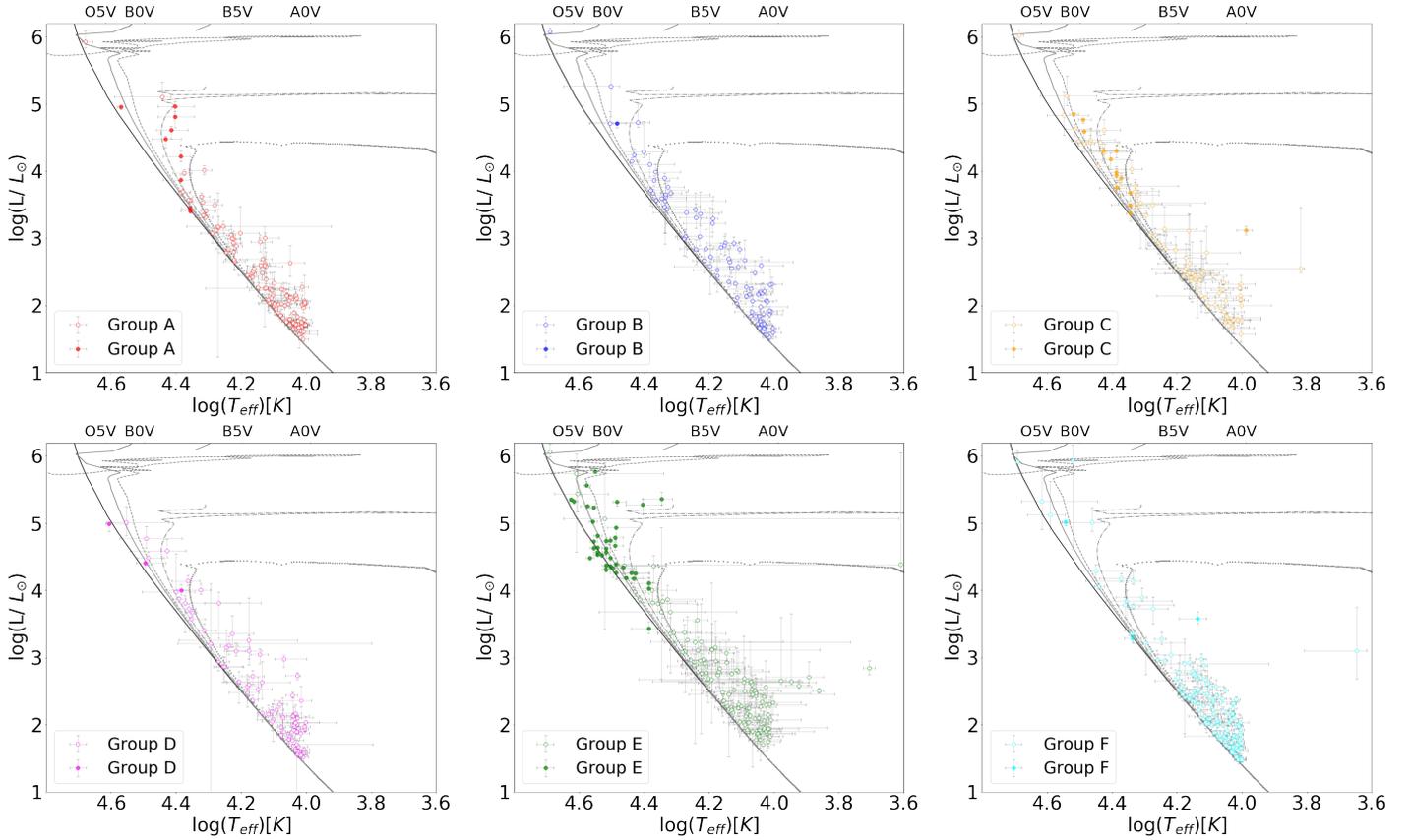}
    \caption{HR diagrams for all the groups. Full coloured symbols correspond to those with spectroscopy while the empty symbols are those without. Isochrones has been drawn from the rotational models in \citet{Ekstrom}. From top to bottom these are the 3.16, 5, 10 and 20 Myr isochrones. Positions of some spectral types have been indicated on the top horizontal axis for clarity. \label{HRdiagramgroups}}
\end{figure*}

HR diagrams for all the groups are shown in Fig. \ref{HRdiagramgroups} using spectroscopic (where available) or SED-fitted (otherwise) effective temperatures and luminosities derived using the fitted distances, extinctions, K-band magnitudes and bolometric corrections (the latter from \citealt{Jordi}).

Estimating the ages of these groups is difficult. Many of the groups include members close to the ZAMS at high luminosities, implying the existence of particularly massive (and therefore young) stars. These would possibly need to be as young as 3--5 Myr, depending on initial rotation rate. There are notable examples of this in groups B, C and E (Cyg OB2). Some of the groups also include members moving away from the main sequence, particularly around the 10 Myr isochrone, implying the presence of stars around this age. Examples of this can be found in groups A, D, E (to a lesser extent) and F.

Using these indicators of age we can place the groups in an approximate age order, with groups B, C and E as the youngest (due to the presence of very luminous stars within them), and groups A, D and F as the oldest (due to the presence of stars at or beyond the 10 Myr isochrone). However, we note that many OB associations, including these, exhibit large age spreads (see \citealt{Wright2020} for examples) and therefore assigning a single age to any association is difficult.

\subsection{Kinematics of the individual groups}
\label{kinematics}

We calculated the median positions and proper motions of each group, along with their distance. The proper motion dispersions were computed using the method from \citet{Ivezic}, again with random gaussian sampling to estimate uncertainties. Velocity dispersions range between 1--3 km s$^{-1}$ as indicated in Table \ref{TabParamGroups}, consistent with other OB associations \citep{Wright2020}.

\begin{figure}
    \centering
    \includegraphics[scale = 0.35]{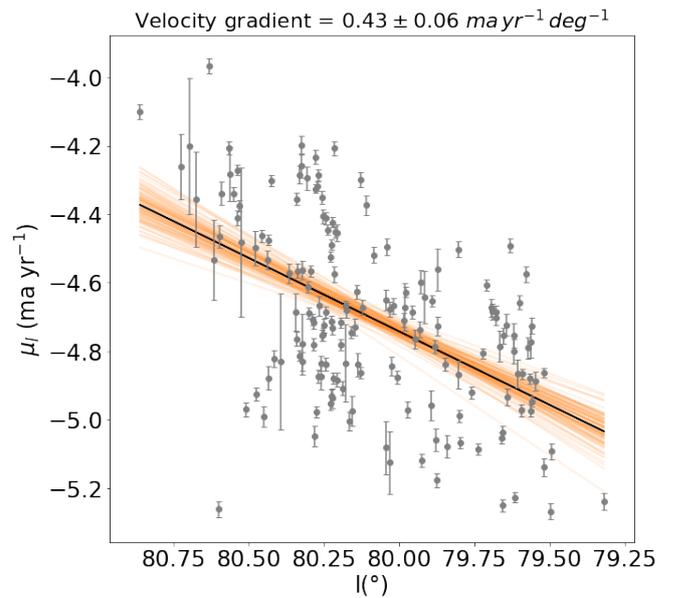}
    \caption{Results of the MCMC fit between position and proper motion for the $l $ direction for group E. 100 random samples have been drawn from the posterior distribution. The black line corresponds to the best-fit value for the velocity gradient which has been indicated on the top of the plot.  \label{Fitpospm}}
\end{figure}

To determine whether the groups are expanding, we search for evidence of velocity gradients in the kinematics of our stars.  To compute the velocity gradients, we fit a linear relationship using an MCMC simulation and the \textit{emcee} package, and obtain velocity gradients for each group (see Table \ref{TabParamGroups}). An example of a fit is shown in Fig. \ref{Fitpospm}. We note that the velocity gradients are generally anisotropic and larger in the $l$ direction compared to the $b$ direction, similar to the pattern of anisotropic expansion observed in other OB associations such as Sco-Cen \citep{WrightMamajek}.

Velocity gradients can then be converted to expansion ages, listed in Table \ref{TabParamGroups}. Expansion ages are based on the assumption that the entire group expanded from a compact region of space, and therefore represent upper limits rather than precise ages. We have not calculated these quantities in Galactic latitude for the groups B, D and F as their velocity gradients are either too small or negative (both of which result in unphysical expansion ages). On the other hand, results for the $l$ direction tend to be closer to the likely ages for these systems. The expansion ages in the $l$ and $b$ directions do not agree for any of the groups except group A, as has been observed in many other OB associations (see e.g., \citealt{Wright2020}).

While these expansion ages should be considered as upper limits, it is reassuring to see that the three groups we identified as being the youngest (B, C and E, see Section \ref{Known}) have the smallest expansion ages, and the three older groups have the largest expansion ages.

\subsection{Large-scale dynamics of the Cygnus associations}
\label{largescale}

In the right-hand panel of Fig. \ref{PmDistribExtended} the proper motion in Galactic longitude is plotted against Galactic longitude. A very clear trend of increasing proper motion is shown as a function of Galactic longitude for the stars in two combinations of associations: groups A, D and F at low longitudes and groups B, C and E at high longitudes. Notably, these two sets correspond to the three young groups and the three old groups (Section \ref{HRs}), hinting at a connection in both dynamics and age. This pattern repeats itself on a scale of five degrees in longitude (from l = 81$^{\circ}$ to 76$^{\circ}$ and from l = 76$^\circ$ to 71$^{\circ}$). and approximately 3 mas yr$^{-1}$ in proper motion. At a rough distance of 1800 pc these equate to $\sim$150 pc and $\sim$25 km s$^{-1}$, respectively. No such pattern is seen for $\mu_b$ as a function of either $l$ or $b$.

\section{Discussion}
\label{discussion}

In this section we discuss our results, how they compare to previous studies and what they imply for our general understanding of OB associations. Our main results are:

\begin{itemize}
    \item Most of the historical OB associations in the Cygnus region lack kinematic coherence and therefore do not appear to be genuine OB associations. We identified six new systems that we compared to the previous divisions. There is strong overlap between our group E and Cyg OB2, a partial overlap between group A and Cyg OB3, but no significant overlap between our other groups and the other historical associations.
    \item We calculated the broad properties of each new association, including estimates of their total stellar mass, quantifying their incompleteness. Group E (Cyg OB2) contains by far the largest number of O-type stars and is the most massive.
    \item We calculated the velocity dispersions of each group and searched for evidence that they are expanding. We found strong evidence of expansion for all of the groups in the $l$ direction and in groups A, C and E in the $b$ direction.
    \item We discovered a correlation between $l$ and $\mu_l$ on large scales across the entire Cygnus region, specifically connecting the three oldest and the three youngest groups, thus showing a connection between age and dynamics. 
\end{itemize}

\subsection{The new Cygnus OB associations}

We have identified 6 association-like groups in Cygnus, each of which has a total mass of between 1500 and 4500 $M_{\odot}$. These groups are remarkably kinematically-coherent (see Fig. \ref{SpaceDistribExtended} and \ref{PmDistribExtended}), especially when compared to the historical OB associations (Section \ref{Known}). Comparing these groups to the historical associations, the best match is between Group E and Cyg OB2 with 64 stars in common, yet our group has a better kinematic coherency.

\citet{Zari2021} provided a map of hot and luminous OBA stars in the local part of the Milky Way and noticed that massive star forming regions and OB associations appeared as over-densities. Yet only Cyg OB2 is clearly prominent in their map, meaning that in spite of their completeness, they were not able to identify some of the groups that we have. Cyg OB2 is the densest and most massive group of the region, while the other groups have a lower density, by approximately an order of magnitude, explaining why \citet{Zari2021} did not detect such groups.

\citet{Berlanas2} divided Cyg OB2 in two substructures along the line of sight, with a main group situated around 1760 pc and a smaller, foreground group at 1350 pc, with group E matching the former one. \citet{Orellana} refined the census of the region by identifying 2767 proper motion members of Cyg OB2 (with a mean distance of $1669 \pm 5$ pc, very close to the median distance of group E in Table \ref{TabParamGroups}). The existence of two separates structures composing Cyg OB2 could explain why our group E does not include some of the old classical members of Cyg OB2 and why our estimate of the total mass of group E is smaller than the mass of Cyg OB2 estimated by \citet{Wright}. Nonetheless, given their close characteristics, such as central coordinates, distance, total stellar mass, age and high number of O-type stars \citep{Wright, Wright2016, Berlanas2, Orellana}, the newly identified stars in group E are likely to be part of Cyg OB2.

\subsection{Expansion}
\label{expansion}

The expansion of OB associations has been assumed since their very first discovery as they were argued to be the expanded remnants of dense star clusters \citep{Amb1947}. \citet{Wright2020} emphasized the two 'extreme' scenarios for the origin of OB associations being hierarchical and clustered star formation. In the latter case, OB associations are formed when star clusters are disrupted by processes such as residual gas expulsion (see e.g. \citealt{Lada2003}), while in the former case multiple structures may form hierarchically and drift apart without a clearly coherent expansion pattern. Which of these two theories can best explain the highly anisotropic expansion patterns now being seen in many OB associations remains to be seen (though see \citealt{Kru} for a possible cause of asymmetric expansion).

We measured a clear expansion trend in all of the identified groups in the $l$ direction along with a similar trend in the $b$ direction for the groups A, C and E. By comparison, \citet{Wright2016} found no clear evidence of expansion in Cyg OB2. \citet{WrightMamajek} also failed to detect expansions in both directions in Sco-Cen, neither did \citet{Ward} for their 18 studied OB associations. Using {\it Gaia} DR2 data, \citet{Melnik2020} only identified expansion in 6 out of their 28 selected OB associations. On the other hand, \citet{CantatGaudin2019} and \citet{Armstrong2020} measured expansion in sub-groups inside the Vela-Puppis region, albeit anisotropic expansion, while \citet{Kounkel2018} made a similar discovery in the substructures of the Orion complex. Our results fall in the second category and share the feature with those studies that the OB association groups have been defined kinematically. The fact that our group E is expanding while \citet{Wright2016} did not detect any expansion in the historical Cyg OB2 illustrates this trend: \citet{Wright2016} included some stars in Cyg OB2 with different kinematics and potentially at a different distance \citep{Berlanas2}.

It is clear that studies that used the historical membership of an OB association have typically failed to find evidence of expansion, while those studies that have redefined the membership using kinematic information have more consistently identified expansion trends.

\subsection{Large-scale kinematics and expansion}

The large-scale kinematics in the right panel of Fig. \ref{PmDistribExtended} reveal two patterns with a length of about five degrees in Galactic longitude and 3 mas yr$^{-1}$ in proper motion, equating to a length of about 150 pc and a velocity of 25 km s$^{-1}$ (at a distance of 1.8 kpc). The direction of Galactic longitude towards the Cygnus region is approximately transverse to the direction of the Cygnus spur (or spiral arm) within the plane of the Galactic disk. A similar kinematic pattern has recently been observed in the Carina arm by \citeauthor{Drew2021} (submitted).

What could produce such a large-scale, modulating kinematic pattern amongst these young stars? We explored whether such a kinematic pattern could emerge due to the projection of Galactic rotation, using the rotation law of \citet{Eilers}, individual distances, and calculating Galactocentric radii for each star. The kinematic signature that results does show a dependence between $l$ and $\mu_l$, but it is smaller in $\mu_l$ by almost an order of magnitude than that observed, and does not exhibit the sawtooth pattern seen in Fig. \ref{PmDistribExtended}. This implies that the observed kinematic pattern originates not from Galactic rotation but from the local dynamics in Cygnus. 

A correlation between distance and velocity in the same direction is often an indication of expansion, which can arise due to feedback. \citet{Chevance} have argued that feedback within H~{\sc II} regions promotes velocities of order $\sim$15 km s$^{-1}$, which would likely occur in opposite directions and thus lead to a kinematic pattern with a magnitude of $\sim$30 km $s^{-1}$, very similar to the 25 km $s^{-1}$ pattern we observe. \citet{Chevance} observe this phenomena in other galaxies on scales of 100--300 pc, again consistent with the scale of 160 pc we observe. This could indicate that the kinematic pattern we are observing was introduced by a previous generation of stars that exerted feedback on the surrounding gas clouds, that then went on to form the stars we have studied here, with those stars inheriting the motion of the gas. We note also that our 6 new OB associations are arranged in this large-scale kinematic pattern in groups of 3 OB associations. From analysis of the HR diagrams of these groups we estimated that the three youngest associations are part of one of these large-scale patterns and the three oldest associations are part of the second pattern. This temporal link lends weight to the idea that a common origin was responsible for these two large-scale structures.

An argument against such an interpretation is that one would naively expect such an expansion signature to be present in the $b$ direction as well as in $l$, yet we do not observe this. This could be because of the restoring force of the Galactic disk inhibiting such expansion, or because the gas clouds expanding in the $b$ direction did not form a new generation of stars, possibly because of the lower-density medium they expanded into that did not trigger further star formation.

Alternatively, the kinematic pattern we observe could be due to Galactic shear forces acting on the primordial giant molecular cloud, with the forming stars inheriting this motion. This interpretation has been argued to explain why the expansion of the Sco-Cen association is primarily in the Galactic $Y$ direction \citep{WrightMamajek}. However, the timescales for Galactic shear to act are $\sim$70~Myr \citep{DobbsPringle2013}, too short to explain the observed kinematics. Further observations and simulations appear necessary to explain this interesting kinematic pattern.

\section{Conclusions}
\label{conclusions}

In this paper we have investigated the existence of the OB associations in Cygnus, concluded that the current associations, on the whole, do not represent kinematically-coherent groups of stars, and identified 6 new groups of OB stars that appear more coherent and are therefore likely to be real OB associations.

We have identified OB stars using an SED fitting process, that exploits photometry and astrometry from large-scale surveys  and evolutionary models. This allowed us to identify 4680 candidate OB stars, including 818 probable O-type stars, with an estimated reliability of $>$90\%. From this sample, and using a new and flexible tool to identify coherent kinematic groups, we identified 6 new OB associations in Cygnus. We compared them with the historical OB associations and found some overlap between Cyg OB2 and Cyg OB3 and our new associations, but very little overlap with any of the other historical associations.

We measured several physical and kinematic properties of these groups and found a good consistency with other OB associations, notably for velocity dispersion, total mass and age. Expansion was identified and measured in the $l$ direction for all the groups and in the $b$ direction for three of them. This result is comparable to other recent studies that have detected expansion in OB associations when they have been identified or characterised using kinematic criteria and contrasting with studies that have used the historical membership of OB associations and typically failed to find signatures of expansion.

This highlights the importance of revisiting the historical OB associations. It leads not only to new divisions with a higher kinematic consistency, but also to better characterization of the systems and their expansion. This method can easily be applied to other regions or OB associations to validate the reality of existing systems or identify new ones.

\section*{Acknowledgements}

The authors would like to thank the anonymous referee for many helpful suggestions that improved the quality of this work and illuminated possible avenues for further work. ALQ acknowledges receipt of an STFC postgraduate studentship. NJW acknowledges receipt of an STFC Ernest Rutherford Fellowship (ref. ST/M005569/1) and a Leverhulme Trust Research Project Grant (RPG-2019-379). The authors would also like to thank Janet Drew and Maria Mongui\'o for discussions on Galactic rotation and its kinematic signature.

This paper makes uses of data processed by the Gaia Data Processing and Analysis Consortium (DPAC, https://www.cosmos.esa.int/web/gaia/dpac/consortium) and obtained by the Gaia mission from the European Space Agency (ESA) (https://www.cosmos.esa.int/gaia), as well as the INT Galactic Plane Survey (IGAPS) from the Isaac Newton Telescope (INT) operated in the Spanish Observatorio del Roque de los Muchachos. Data were also based on the Two Micron All Star Survey, which is a combined mission of the Infrared Processing and Analysis Center/California Institute of Technology and the University of Massachusetts, along with The UKIDSS Galactic Plane Survey (GPS), a survey carried out by the UKIDSS consortium with the Wide Field Camera performing on the United Kingdom Infrared Telescope.

This work also benefited from the use of \textit{TOPCAT} \citep{Topcat}, Astropy \citep{Astropy} and the Vizier and SIMBAD database, both operated at CDS, Strasbourg, France.

\section*{Data availability}
The data underlying this article will be shared on reasonable request to the corresponding author.



\bibliographystyle{mnras}
\bibliography{Bibliography}


\appendix

\section{Comparison between our groups}
\label{compgroups}

\begin{figure*}
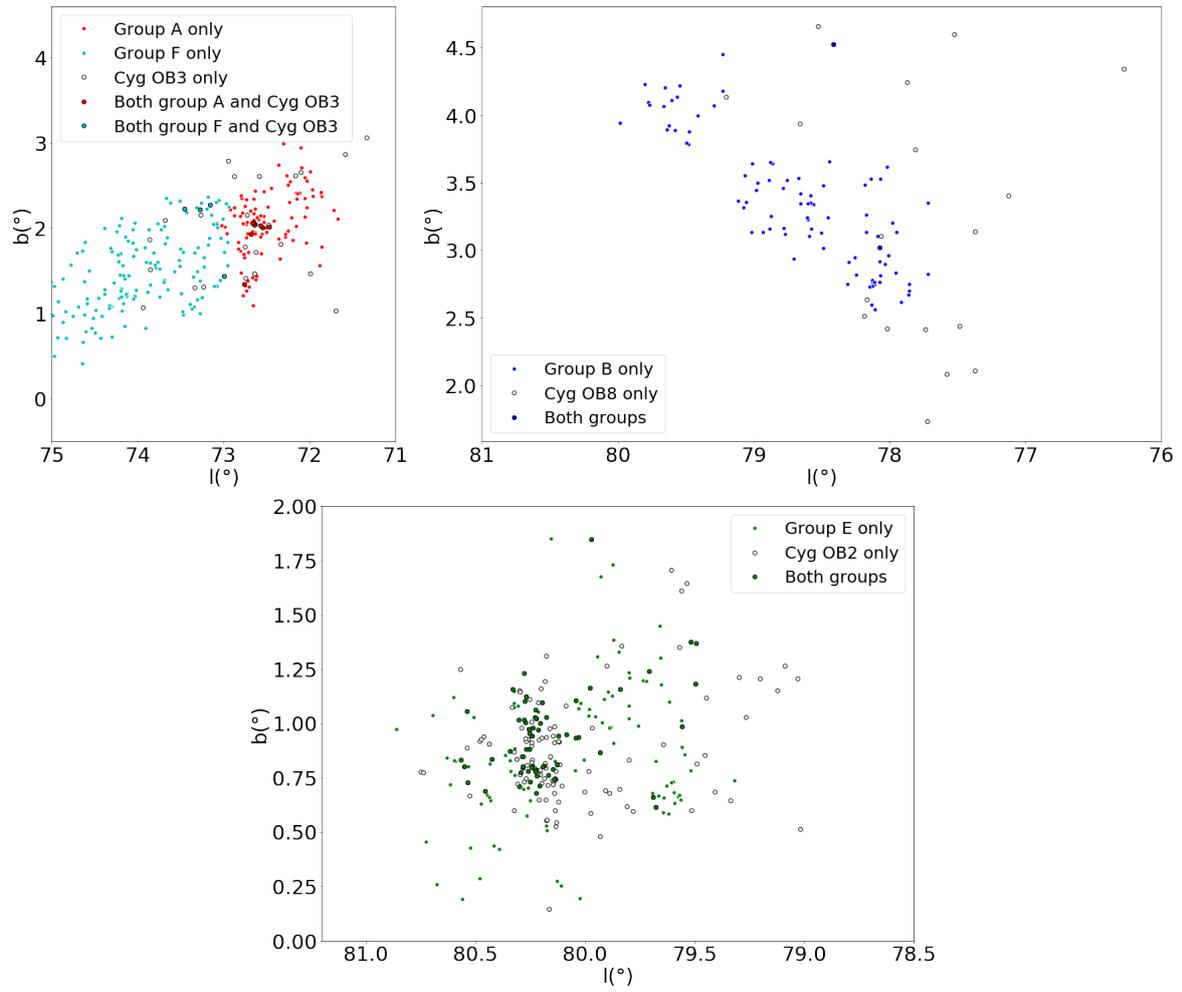

    \centering
    \includegraphics[scale = 0.25]{GroupafCygOB3.pdf}
    \includegraphics[scale = 0.25]{GroupbCygOB8.pdf}
    \includegraphics[scale = 0.25]{GroupeCygOB2.pdf}
    \caption{Comparison between our new groups and the historical associations Cyg OB2, OB3 and OB8.    \label{CompHist}}
\end{figure*}

Figures \ref{CompHist} and \ref{CompDBSCAN} show a visual comparison between our new groups and, respectively, the historical OB associations in this part of Cygnus and the groups identified using DBSCAN. See Sections \ref{HistCompSection} and \ref{DBSCANComp} for a discussion of the overlap in membership between our groups and these groups.

\begin{figure*}
    \centering
    \includegraphics[scale=0.25]{Groupa6.pdf}
    \includegraphics[scale =0.25]{Groupb5.pdf}
    \includegraphics[scale = 0.25]{Groupcd1.pdf}
    \includegraphics[scale = 0.25]{Groupcd3.pdf}
    \includegraphics[scale = 0.25]{Groupe2.pdf}
    \caption{Comparison between our new groups and the groups identified by DBSCAN.    \label{CompDBSCAN}}
\end{figure*}

\bsp	
\label{lastpage}
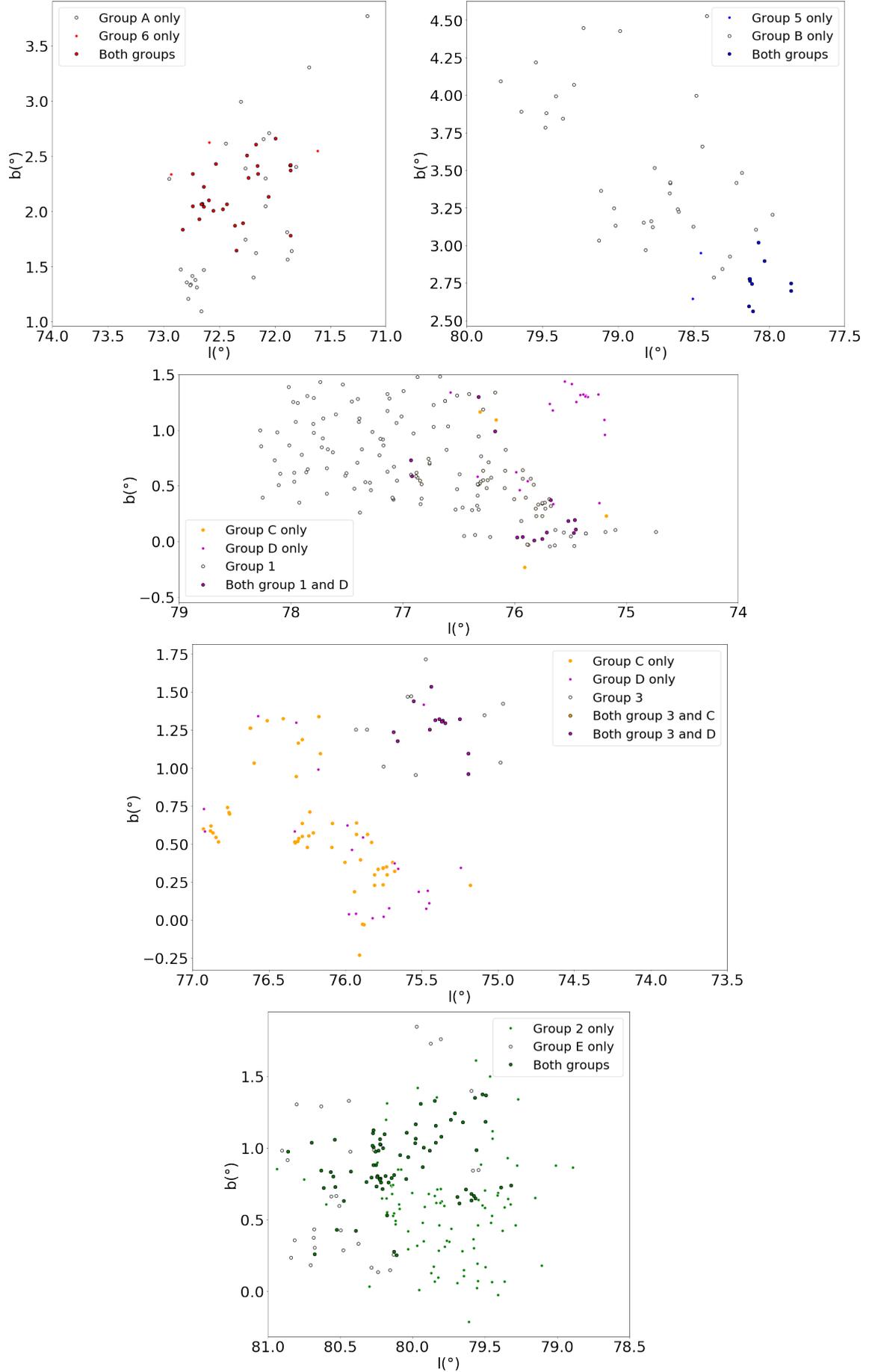
\end{document}